\documentclass[aps,preprint,amsmath,amssymb]{revtex4}
\usepackage{graphicx}
\usepackage{float}
\begin{document}

\title{Model-independent sensitivity estimates for the electromagnetic dipole moments of the top quark at the Large Hadron Collider and beyond}

\author{M. Koksal}
\email[]{mkoksal@cumhuriyet.edu.tr} \affiliation{Department of
Optical Engineering, Sivas Cumhuriyet University, 58140, Sivas, Turkey}

\begin{abstract}

As the heaviest known fundamental particle, the top quark ensures testing of the Standard Model and occupies a significant role in a lot of theories of new physics beyond the Standard Model. Up to now, the top quark has been only generated at the Tevatron and the Large Hadron Collider. However, one of the most important tasks of the physics program at the Large Hadron Collider is the investigation of the anomalous top quark interactions. In addition, the production of single top quarks, though rarer than production in pairs, presents us a different way to study the top quark produced via the electroweak interaction. This makes single top quark production an important signature for studying the properties of the top quark. Hence, we examine the anomalous $t\bar{t}\gamma$ interactions to investigate limits on the anomalous $a_A$ and $a_V$ couplings through single top quark production of the process $pp\rightarrow p \gamma^{*} p \rightarrow p t W X$ at the Large Hadron Collider, the High Luminosity Large Hadron Collider and High Energy Large Hadron Collider. The best limits obtained on the anomalous $a_A$ and $a_V$ couplings through the subprocess $\gamma^{*} b \rightarrow W t$ at the High Energy Large Hadron Collider with $L_{int}=15$ ab$^{-1}$ at $68\%$ Confidence Level are found to be $|a_{A}|=0.0590$ and $a_{V}=[-0.3202;0.0108]$. Therefore, we understand that $\gamma^{*} p$ collisions at the High Energy Large Hadron Collider improves the sensitivity limits of the anomalous coupling parameters according to previous studies.

\end{abstract}

\maketitle

\section{Introduction}

Particle physics experiments carried out in the last decades indicate that the Standard Model (SM) is consistent in a low energy regime. Nevertheless, there are still some questions to be answered. It is expected that these questions will be answered by experimental studies that are planned to be done now and in the future. Besides, it is thought to be embedded in a more fundamental theory where its effects can be defined at a higher energy regime.

In the SM, the top quark is the heaviest elementary particle that is observed up to today. Since the mass of the top quark is close to the electroweak scale, making it one of the most interesting particles in the SM. For this reason, the top quark canalizes us to study its couplings such as $t\bar{t}\gamma$, $t\bar{t}Z$, $t\bar{t}g$, $t\bar{t}H$, $Wtb$, $tqZ$ and $tq\gamma$ ($q=u,c$). However, some of the properties of the top quark are still poorly constrained such as the magnetic and electric dipole moments and the chromomagnetic and chromoelectric dipole moments. Accordingly, important new insights on the properties of the top quark will be one of the tasks of the Large Hadron Collider (LHC). Therefore, measuring the characteristics of the top quark could give us an opportunity to understand the electroweak sector and the new physics beyond SM.

CP violation was first observed in 1964 during measurements of the neutral kaon decays \cite{cp1}. CP violation in the context of the SM has been described by the complex couplings in the CKM matrix of the quark sector \cite{cp2}. On the other hand, knowledge coming from the SM is not enough to have deeper understanding of the origin of CP violation. Because the amount of baryons in the universe predicted using the CKM mechanism gives value much lower than the value estimated from the experiment. However, the effect obtained from this study was very small but was highly important because it proved that matter and antimatter are intrinsically different. Therefore, it is essential to research new sources of CP violation \cite{yam,yam1,yam2}. For this reason, the measurement of large amounts of CP violation in the top quark events can be an evidence of physics beyond the SM. Investigating of physics beyond the SM, the magnetic and electric dipole moments of the top quark can be analyzed. The magnetic dipole moment of the top quark is obtained from one-loop level perturbations in the SM, while the electric dipole moment that is determined as a source of CP violation \cite{cp3,cp4} is achieved from three-loop level perturbations. Hence, there can mutate the top quark-photon vertices through CP-symmetric and CP-asymmetric anomalous form factors to explore the non-SM effects in the top quark production reactions.

The SM estimation for the electric dipole moment of the top quark is very small. For this reason, it can be an extremely attractive investigation of physics beyond the SM. Nevertheless, the magnetic dipole moment of the top quark in the SM is not far from the future collider experiments. The values of the magnetic and electric dipole moment of the top quark in the SM are $a_{A}<1.75 \times 10^{-14}$ and $a_{V}=0.013$, respectively \cite{sm,sm1}. Particularly, searches at $pp$ colliders such as the Tevatron and the LHC were suggested to probe the magnetic and electric dipole moments of the top quark in measurements of the reactions $p \bar{p}\rightarrow t\bar{t}\gamma$ \cite{1}, $p p\rightarrow t j\gamma$ \cite{2}, $p p\rightarrow p \gamma^{*} \gamma^{*} p\rightarrow p t \bar{t} p$ \cite{4} and $p p\rightarrow p \gamma^{*} p\rightarrow t \bar{t} p X$ \cite{mur}. Studies on sensitivity of the future $e^{-}e^{+}$ linear colliders and their operating modes of $e \gamma$, $e \gamma^{*}$, $\gamma \gamma$ and $\gamma^{*} \gamma^{*}$ to constrain the electromagnetic dipole moments were analyzed through the reactions $e^{-}e^{+}\rightarrow t \bar{t}$ \cite{5}, $\gamma e\rightarrow \bar{t} b \nu_{e}$ \cite{6}, $e^{-}e^{+}\rightarrow e^{-}\gamma^{*} e^{+} \rightarrow \bar{t} b \nu_{e} e^{+}$ \cite{6}, $\gamma \gamma \rightarrow t \bar{t}$ \cite{7} and $e^{-}e^{+}\rightarrow e^{-}\gamma^{*} \gamma^{*} e^{+} \rightarrow e^{-}t \bar{t}e^{+}$ \cite{7}. In addition, the reactions $ep\rightarrow \bar{t} \nu_{e} \gamma$ \cite{8}, $ep\rightarrow e \gamma^{*} p\rightarrow \bar{t} \nu_{e} b p$ \cite{tur} and $ep\rightarrow e \gamma^{*} \gamma^{*} p\rightarrow e t \bar{t} p$ \cite{10} on the future $ep$ colliders are phenomenologically investigated. All limits obtained in the above investigations are shown in Table I. However, there are other studies related to the anomalous $t\bar{t}\gamma$ vertices that can observe the magnetic or electric dipole moments of the top quark in the literature \cite{9,20,21,22,23,24,3}.

\begin{table}[!ht]
\caption{Sensitivity limits on the magnetic and electric dipole moments of the top quark through different processes at $pp$, $e^{-} e^{+}$ and $ep$ colliders}
\begin{center}
\begin{tabular}{|c|c|c|}
\hline\hline
{\bf Processes}  &    {\bf $a_{V}$}  &    {\bf $a_{A}$}  \\
\hline
$p p\rightarrow t\bar{t}\gamma$ \cite{1}           &   $ (-0.200, 0.200)$ & $(-0.100, 0.100) $ \\
\hline
$p p\rightarrow t j\gamma$ \cite{2}           &   $ (-0.110, 0.110)$ & $ (-0.090, 0.090) $   \\
\hline
 $pp \to p\gamma^*\gamma^*p\to pt\bar t p $ \cite{4}      &   $ (-0.458, 0.016)$ & $ (-0.081, 0.081) $  \\
\hline
$p p\rightarrow p \gamma^{*} p\rightarrow t \bar{t} p X$ \cite{mur} &   $ (-0.9953, 0.0003)$ & $ (-0.020, 0.020) $ \\
\hline
$e^+e^- \to t\bar t$ \cite{5}         &   $ (-0.002, 0.002)$ & $ (-0.001, 0.001) $   \\
\hline
$\gamma e\rightarrow \bar{t} b \nu_{e}$ \cite{6}         &   $ (-0.027, 0.036)$ & $ (-0.031, 0.031) $ \\
\hline
$e^{-}e^{+}\rightarrow e^{-}\gamma^{*} e^{+} \rightarrow \bar{t} b \nu_{e} e^{+}$ \cite{6}         &   $ (-0.054, 0.092)$ & $ (-0.071, 0.071) $  \\
\hline
$\gamma \gamma \rightarrow t \bar{t}$ \cite{7}         &   $ (-0.023, 0.032)$ & $ (-0.027, 0.027) $ \\
\hline
$e^{-}e^{+}\rightarrow e^{-}\gamma^{*} \gamma^{*} e^{+} \rightarrow e^{-}t \bar{t}e^{+}$ \cite{7}         &   $ (-0.054, 0.092)$ & $ (-0.071, 0.071) $ \\
\hline
$ep\rightarrow \bar{t} \nu_{e} \gamma$ \cite{8} &   $ (-0.269, 0.251)$ & $ (-0.259, 0.259) $  \\
\hline
$ep\rightarrow e \gamma^{*} p\rightarrow \bar{t} \nu_{e} b p$ \cite{tur} &   $ (-0.077, 0.073)$ & $ (-0.075, 0.075) $    \\
\hline
$ep\rightarrow e \gamma^{*} \gamma^{*} p\rightarrow e t \bar{t} p$ \cite{10} &   $ ( -0.468, 0.0177)$ & $ (-0.088, 0.088) $    \\
\hline
\end{tabular}
\end{center}
\end{table}

Presently, phase 2 of the LHC has finished and has closed for an upgrade during the period 2019-2020 years. In future times, the LHC is going to activate at $\sqrt{s}=14$ TeV with $L_{int}=300$ fb$^{-1}$ between 2021-2023 years. Besides, the High Luminosity Large Hadron Collider (HL-LHC) project aims to crank up the performance of the LHC in order to increase the potential for discoveries after 2025. The HL-LHC is expected to operate for ten years until 2036 year. After this period, it is guessed that each detector will collect nearly 3000 fb$^{-1}$ data \cite{HL}. Other colliders other than HL-LHC are also discussed. Also, in the frame of the FCC study, there are designing a $\sqrt{s}=27$ TeV hadron collider in the LHC tunnel, called the High Energy Large Hadron Collider (HE-LHC). It will collect a dataset corresponding to an integrated luminosity of 10-15 ab$^{-1}$ \cite{HE}.

In many experiments carried out at LHC, new physics studies have been tested via $pp$ processes that comprise research through subprocesses of gluon-gluon, quark-quark and quark-gluon collisions. Nevertheless, precise measurements may not be possible due to proton remnants occurring after these collisions and it may be difficult to detect the signs which may arise from the new physics. In addition to this, $\gamma^{*} \gamma^{*}$ and $\gamma^{*} p$ collisions that provide a more clean environment with respect to $pp$ collisions are much less investigated. Both of the incoming protons in $\gamma^{*} \gamma^{*}$ collisions remain intact and do not dissociate into partons. Also, while only one of the incoming protons in $\gamma^{*} p$ collisions dissociates into partons, the other proton remains intact. The cleanest background between these two collisions is $\gamma^{*} \gamma^{*}$. $\gamma^{*} p$ collisions have more background than $\gamma^{*} \gamma^{*}$ processes. Moreover, $\gamma^{*} p$ collisions have effective luminosity and much higher energy compared to $\gamma^{*} \gamma^{*}$ collisions. This phenomenon may be important due to the high energy dependencies of the total cross section of the process including the new physics beyond the SM. Therefore, $\gamma^{*} p$ collisions are expected to have a high sensitivity to the new physics parameters.

Specially, the high cross section obtained in $\gamma^{*} p$ collisions is very important for examining the properties of the top quarks produced in the final state. The production of a $W$ boson and a top quark is very valuable, as it investigates the $|V_{tb}|$ element of the Cabibbo-Kobayashi-Maskawa matrix and the anomalous $Wtb$ couplings \cite{ckm,ckm2}. The anomalous $tq\gamma$ couplings with single top quark production are examined by Refs. \cite{ahe,ah1}. Also, the measurement of charged top-pion $\pi_{t}^{\pm}$ production associated with a top quark is presented \cite{ahe1}. Events related to $t \bar{t}$ pair could be used to study parameters like the charge or mass of the top quark \cite{ahe2}. Also, $t \bar{t}$ pair production to investigate the sensitivity on the magnetic and electric dipole moments of the top quark is discussed \cite{mur}.

Photons emitted from the proton beams in $\gamma^{*} \gamma^{*}$ and $\gamma^{*} p$ collisions can be described in the framework of the Equivalent Photon Approximation (EPA) \cite{epa,epa1,epa2}. Photons in this approach are almost real because they have low virtuality. The EPA has many advantages such as providing the skill to reach crude numerical predictions through simple formulae. Nevertheless, this approach can usually simplify the experimental analysis because it provides an occasion one to directly get a rough cross-section for $\gamma^{*} p \rightarrow X$ subprocess via the investigation of the process $pp \rightarrow p \gamma^{*} p \rightarrow X p$ where $X$ symbolizes objects produced in the final state. New physics studies beyond the SM are generally examined by using the EPA \cite{s1,s2,s3,s4,s5,s6,s7,s8,s9,s10,s11,s12,s13,s15,s16,s18,s19,s20,s21,s22,sa1,sa2,sa3,sa4,sa5,sa6,sa7,sa10,sa11,sa12}.

\section{Single top quark production in $\gamma^{*} p$ collisions}

\subsection{Effective Lagrangian of $t\bar t \gamma$ couplings}

Anomalous $t\bar t \gamma$ couplings can be examined in a model-independent way  by means of effective Lagrangian approach. In this study, we consider the following effective Lagrangian defining the anomalous $t\bar t \gamma$ couplings \cite{Kamenik,Baur,Aguilar1}

\begin{equation}
{\cal L}_{t\bar t\gamma}=-e Q_t\bar t \Gamma^\mu_{ t\bar t  \gamma} t A_\mu
\end{equation}
where $e$ is the proton charge, $Q_t$ is the top quark electric charge, $A_\mu$ is the photon gauge field. $\Gamma^\mu_{t\bar t \gamma}$ is given as follows

\begin{equation}
\Gamma^\mu_{t\bar t\gamma}= \gamma^\mu + \frac{i}{2m_t}(a_V + i a_A\gamma_5)q_\nu \sigma^{\mu\nu}
\end{equation}

\noindent where $m_t$ represents the mass of top quark, $q_\nu$ defines the photon four-momentum, $\gamma_5 q_\nu$ term with $\sigma^{\mu\nu}$ breaks the CP symmetry. Therefore, $a_A$ parameter shows the strength of a possible CP violation process, which may be caused by new physics beyond the SM. Real $a_V$ and $a_A$ parameters are non-SM couplings and interested in the anomalous magnetic and the electric dipole moments of the top quark, respectively.

\subsection{Theoretical Calculations}

A schematic diagram for the process $pp\rightarrow p \gamma^{*} p \rightarrow p t W X$ is given in Fig. 1. Here, photon emitted from the incoming proton beam can collide the other proton and produce $\gamma^{*} p$ collision.

\begin{figure} [!htb]
\begin{center}
\includegraphics [width=8cm,height=5cm]{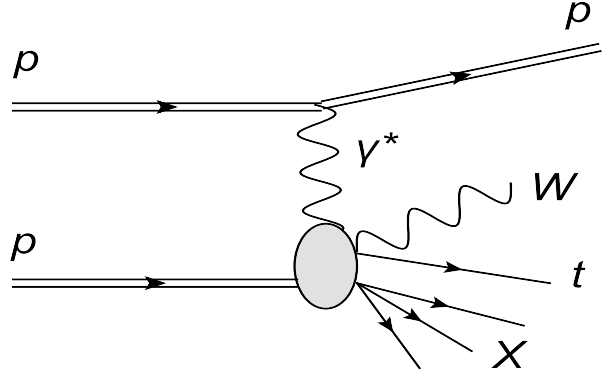}
\caption{Diagram for the process $pp\rightarrow p \gamma^{*} p \rightarrow p t W X$.
\label{fig1}}
\end{center}
\end{figure}

In this collision, the photon spectrum in terms of virtuality $Q^{2}$ and energy $E_{\gamma}$ is given by \cite{epa}

\begin{eqnarray}
\frac{d N_{\gamma}}{d E_{\gamma} d Q^{2}}=\frac{\alpha}{\pi}\frac{1}{E_{\gamma}Q^{2}}[(1-\frac{E_{\gamma}}{E})(1-\frac{Q_{min}^{2}}{Q^{2}})F_{E}+\frac{E^{2}_{\gamma}}{2E^{2}}F_{M}]
\end{eqnarray}
where
\begin{eqnarray}
Q_{min}^{2}=\frac{m_{p}^{2}E^{2}_{\gamma}}{E(E-E_{\gamma})},
\end{eqnarray}

\begin{eqnarray}
F_{E}=\frac{4 m_{p}^{2}G^{2}_{E}+Q^{2}G_{M}^{2}}{4 m_{p}^{2}+Q^{2}},
\end{eqnarray}

\begin{eqnarray}
G_{E}^{2}=\frac{G_{M}^{2}}{\mu_{p}^{2}}=(1+\frac{Q^{2}}{Q_{0}^{2}})^{-4},
\end{eqnarray}

\begin{eqnarray}
F_{M}=G_{M}^{2},
\end{eqnarray}

\begin{eqnarray}
Q_{0}^{2}=0.71 GeV^{2}
\end{eqnarray}
where $m_{p}$ is the proton mass, $E$ is the energy of the incoming proton beam, the magnetic moment of the proton is
$\mu_{p}^{2}=7.78$, $F_{M}$ and $F_{E}$ are functions of the magnetic and electric form factors, respectively.

After integration over $Q^{2}$, the photon spectrum is derived as follows

\begin{eqnarray}
\frac{d N_{\gamma}}{d E_{\gamma}}=\frac{\alpha}{\pi E_{\gamma}}\{[1-\frac{E_{\gamma}}{E}][\varphi(\frac{Q_{max}^{2}}{Q_{0}^{2}})-\varphi(\frac{Q_{min}^{2}}{Q_{0}^{2}})]
\end{eqnarray}
where the function $\varphi$ is shown by

\begin{eqnarray}
\varphi(\theta)=&&(1+ay)\left[-\textit{In}(1+\frac{1}{\theta})+\sum_{k=1}^{3}\frac{1}{k(1+\theta)^{k}}\right]+\frac{y(1-b)}{4\theta(1+\theta)^{3}} \nonumber \\
&& +c(1+\frac{y}{4})\left[\textit{In}\left(\frac{1-b+\theta}{1+\theta}\right)+\sum_{k=1}^{3}\frac{b^{k}}{k(1+\theta)^{k}}\right]. \nonumber \\
\end{eqnarray}

$y$, $a$, $b$ and $c$ parameters in above equation are given as follows

\begin{eqnarray}
y=\frac{E_{\gamma}^{2}}{E(E-E_{\gamma})},
\end{eqnarray}
\begin{eqnarray}
a=\frac{1+\mu_{p}^{2}}{4}+\frac{4m_{p}^{2}}{Q_{0}^{2}}\approx 7.16,
\end{eqnarray}
\begin{eqnarray}
b=1-\frac{4m_{p}^{2}}{Q_{0}^{2}}\approx -3.96,
\end{eqnarray}
\begin{eqnarray}
c=\frac{\mu_{p}^{2}-1}{b^{4}}\approx 0.028.
\end{eqnarray}

The subprocess $\gamma^{*} b \rightarrow t W$ that has 3 Feynman diagrams is given in Fig. 2. Only one of the diagrams contains the anomalous $t\bar{t}\gamma$ coupling while the others show contributions arising from the SM. In this study, all numerical calculations have been calculated using the computer package CalcHEP \cite{calc} by embedding $t\bar{t}\gamma$ vertex in Eq. (2). In addition, we have used CTEQ6L1 \cite{cte} for the parton distribution function of the proton.

\begin{figure} [!htb]
\includegraphics [width=10cm,height=7cm]{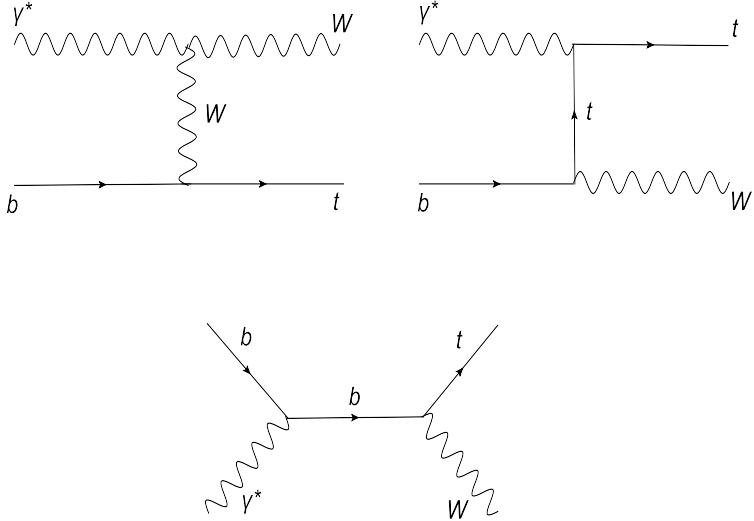}
\caption{Feynman diagrams for the subprocess $\gamma^{*} b \rightarrow t W$
\label{fig2}}
\end{figure}

The total cross sections of the process $pp\rightarrow p \gamma^{*} p \rightarrow p t W X$ as functions of the anomalous $a_A$ and $a_V$ couplings at the LHC, the HL-LHC and the HE-LHC are shown in Figs. 3-4. Here, one of the anomalous couplings is non-zero at any time, while the other coupling is fixed to zero.
The deviation from the SM value of the examined process at the HE-LHC option are greater than the LHC and the HL-LHC options as one can expect due to higher center-of-mass energy. Therefore, the limits obtained on these couplings at the HE-LHC option are expected to be more restrictive than the limits at the LHC and the HL-LHC options. Actually, the anomalous $a_A$ and $a_V$ couplings have different CP properties. Thus, since the cross sections have only even powers of the anomalous $a_A$ coupling, a nonzero value of $a_A$ coupling allows a constructive effect on the total cross section. Nevertheless, the cross sections for $a_V$ coupling involve both odd and even powers. Thus, $a_V$ coupling has a partially destructive effect on the total cross section.

\begin{figure} [!htb]
\includegraphics{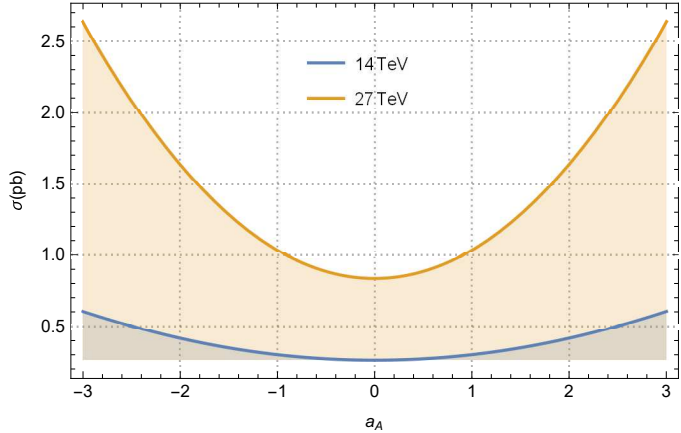}
\caption{The cross sections for the process $pp\rightarrow p \gamma^{*} p \rightarrow p t W X$ X including anomalous $t\bar{t}\gamma$ interactions with
respect to $a_{A}$ coupling.
\label{fig3}}
\end{figure}

\begin{figure} [!htb]
\includegraphics{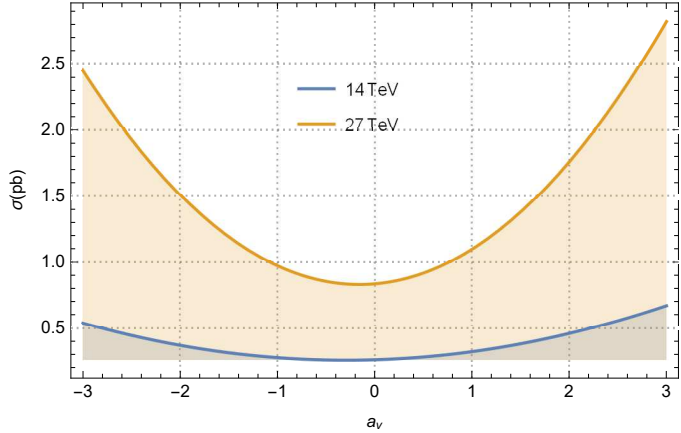}
\caption{Same as in Fig. 3, but for $a_{V}$ coupling.
\label{fig4}}
\end{figure}

We study the total cross section of the process $pp\rightarrow p \gamma^{*} p \rightarrow p t W X$ as a function of the anomalous couplings for the center-of-mass energies of 14 and 27 TeV as shown in Figs. 5-6. The results show a clear dependence of the total cross section of the process according to the anomalous couplings, as well as with the center-of-mass energies.

\begin{figure}
\includegraphics{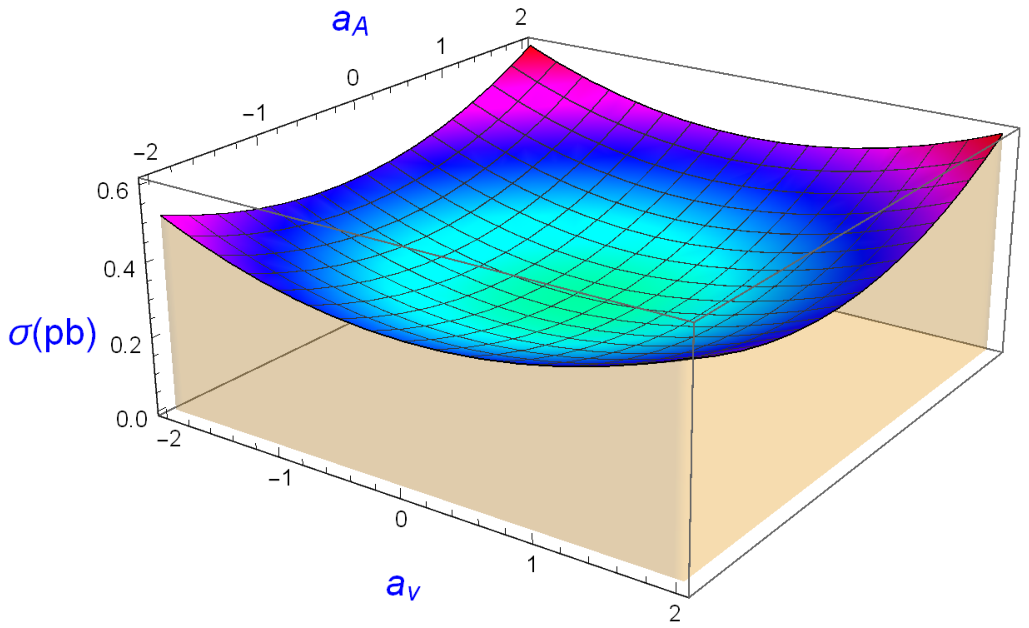}
\caption{The cross section of the process $pp\rightarrow p \gamma^{*} p \rightarrow p t W X$ depending on the anomalous $a_{A}$ and $a_{V}$ couplings at $\sqrt{s}=14$ TeV.
\label{fig5}}
\end{figure}

\begin{figure}
\includegraphics{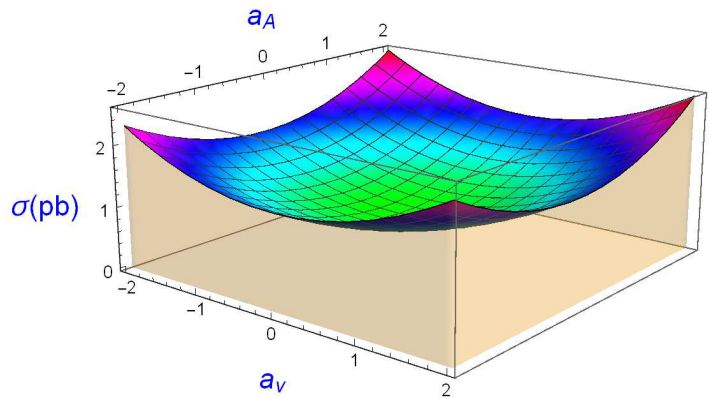}
\caption{Same as in Fig. 5, but for $\sqrt{s}=27$ TeV.
\label{fig6}}
\end{figure}

\section{LIMITS ON THE TOP QUARK'S MAGNETIC AND ELECTRIC DIPOLE MOMENTS AT THE LHC, HL-LHC AND HE-LHC}

In order to investigate sensitivity of the anomalous $a_A$ and $a_V$ couplings through the process $pp\rightarrow p \gamma^{*} p \rightarrow p t W X$
process, we take into account $\chi^{2}$ analysis

\begin{eqnarray}
\chi^{2}=\left(\frac{\sigma_{SM}-\sigma_{NP}}{\sigma_{SM}\delta}\right)^{2}.
\end{eqnarray}
Here, $\sigma_{SM}$ and $\sigma_{NP}$ are the SM cross section and the total cross section containing contributions from the SM and new
physics beyond the SM, respectively. $\delta=\frac{1}{\sqrt{\delta_{stat}^{2}+\delta_{sys}^{2}}}$; $\delta_{sys}$ is systematic uncertainty, $\delta_{stat}=\frac{1}{\sqrt{N_{SM}}}$ represents the statistical error, $N_{SM}=L_{int}\times BR (t \rightarrow W b \rightarrow q q' b) \times BR (W\rightarrow \ell \nu_{\ell}) \times \sigma_{SM}\times b_{tag}$; $L_{int}$ symbolizes the integrated luminosity and $b_{tag}$ tagging efficiency is 0.8. The top quark decays nearly $100\%$ to $b$ quark and $W$ boson. In our calculations, we assume that $W$ bosons in the final state decay semi-leptonically. Hence, we take into account that the branching ratios for $W$ bosons are $BR (W\rightarrow \ell \nu_{\ell})=0.213$ for light leptonic decays ($\ell=e,\mu$) and $BR (W\rightarrow q q')=0.674$ for hadronic decays.

We also include the effects of systematic uncertainties on the limits. For this, a measurement is presented of the associated production of a single top quark and a $W$ boson in $pp$ collisions at $\sqrt{s}=$ 13 TeV by the CMS Collaboration at the CERN LHC \cite{top}. The total uncertainty giving the measurement of the $tW$ production cross section in that study is approximately given $10\%$. There are some researches that consider systematic uncertainties to probe the magnetic and electric dipole moments of the top quark. The processes $\gamma \gamma \rightarrow t \bar{t}$ and $e^{-}e^{+}\rightarrow e^{-}\gamma^{*} \gamma^{*} e^{+} \rightarrow e^{-}t \bar{t}e^{+}$  with systematic uncertainties of $0, 5, 10\%$ are performed in Ref. \cite{7}. In Ref. \cite{9}, the processes $\gamma e\rightarrow \bar{t} b \nu_{e}$, $e^{-}e^{+}\rightarrow e^{-}\gamma^{*} e^{+} \rightarrow \bar{t} b \nu_{e} e^{+}$, $ep\rightarrow e \gamma^{*} p\rightarrow \bar{t} \nu_{e} b p$ are studied from $0\%$ to $5\%$ with systematic uncertainties. In Ref. \cite{20}, a $10\%$ total uncertainty for measurements of the process $\gamma e \rightarrow t \bar{t}$ is considered. For the cross section of the process $pp\rightarrow p \gamma^{*} p \rightarrow p t \bar{t} X$, systematic uncertainties are calculated up to $5\%$ \cite{mur}. Therefore, we present numerical results taking into account systematic uncertainties of $0, 5, 10\%$.

For $68\%$, $90\%$ and $95\%$ Confidence Levels, Tables II-X represent the limits obtained the anomalous $a_{A}$ and $a_{V}$ couplings at the LHC with an integrated luminosity of 300 fb$^{-1}$ with at the HL-LHC with an integrated luminosity of $3$ ab$^{-1}$ and the HE-LHC with an integrated luminosity of $15$ ab$^{-1}$ containing without and with systematic errors $5, 10\%$, respectively. We observe from these tables that limits on the anomalous $a_{A}$ and $a_{V}$ couplings are improved for increasing integrated luminosities and center-of-mass energies. As expected, the best limits on the anomalous couplings between the LHC, the HL-LHC and the HE-LHC are obtained from the HE-LHC option.

As shown in these tables, the best obtained limits on the anomalous $a_{A}$ and $a_{V}$ couplings are given in Table VIII. These are $|a_{A}|=0.0590$ and $-0.3202<a_{V}<0.0108$. Comparing with the anomalous $a_{A}$ coupling derived in the literature, our limit on this coupling is better than those reported in Refs. [10],[11],[12],[17],[18],[19]. In tables, the limits with increasing $\delta_{sys}$ values at the LHC, the HL-LHC and the HE-LHC are nearly unchanged with respect to the values of high center-of-mass energy and high luminosity. The reason of this situation is $\delta_{stat}$ which is much smaller than $\delta_{sys}$.

\begin{table} [!htb]
\caption{The limits obtained at $68\%$ Confidence Level on the electromagnetic dipole moments of the top quark via the process $pp\rightarrow p \gamma^{*} p \rightarrow p t W X$ at the LHC.}
\begin{ruledtabular}
\begin{tabular}{cccc}
Luminosity($fb^{-1}$)&$\delta_{sys}$&$|a_{A}|$ & $a_{V}$  \\
\hline
$10$&$0\%$&$0.6354$ &$ (-0.9907;0.4081)$ \\
$10$&$5\%$&$0.7320$ &$ (-1.0797;0.4970)$ \\
$10$&$10\%$&$0.9012$ &$ (-1.2391;0.6565)$ \\
\hline
$50$&$0\%$&$0.4249$ &$ (-0.8067;0.2241)$ \\
$50$&$5\%$&$0.6292$ &$ (-0.9851;0.4025)$ \\
$50$&$10\%$&$0.8529$ &$ (-1.1933;0.6106)$ \\
\hline
$100$&$0\%$&$0.3573$ &$ (-0.7525;0.1699)$ \\
$100$&$5\%$&$0.6122$ &$ (-0.9697;0.3871)$ \\
$100$&$10\%$&$0.8463$ &$ (-1.1870;0.6043)$ \\
\hline
$300$&$0\%$&$0.2715$ &$ (-0.6896;0.1070)$ \\
$300$&$5\%$&$0.6000$ &$ (-0.9587;0.3761)$ \\
$300$&$10\%$&$0.8417$ &$ (-1.1827;0.6000)$ \\
\hline
\end{tabular}
\end{ruledtabular}
\end{table}

\begin{table} [!htb]
\caption{Same as in Table II, but for $90\%$ Confidence Level}
\begin{ruledtabular}
\begin{tabular}{cccc}
Luminosity($fb^{-1}$)&$\delta_{sys}$&$|a_{A}|$ & $a_{V}$  \\
\hline
$10$&$0\%$&$0.8152$ &$ (-1.1576;0.5750)$ \\
$10$&$3\%$&$0.9392$ &$ (-1.2754;0.6927)$ \\
$10$&$5\%$&$1.1563$ &$ (-1.4846;0.9020)$ \\
\hline
$50$&$0\%$&$0.5451$ &$ (-0.9098;0.3271)$ \\
$50$&$3\%$&$0.8073$ &$ (-1.1502;0.5675)$ \\
$50$&$5\%$&$1.0943$ &$ (-1.4246;0.8420)$ \\
\hline
$100$&$0\%$&$0.4584$ &$ (-0.8348;0.2521)$ \\
$100$&$3\%$&$0.7854$ &$ (-1.1296;0.5470)$ \\
$100$&$5\%$&$1.0858$ &$ (-1.4164;0.8337)$ \\
\hline
$300$&$0\%$&$0.3483$ &$ (-0.7456;0.1629)$ \\
$300$&$3\%$&$0.7698$ &$ (-1.1150;0.5323)$ \\
$300$&$5\%$&$1.0800$ &$ (-1.4108;0.8281)$ \\
\hline
\end{tabular}
\end{ruledtabular}
\end{table}

\begin{table} [!htb]
\caption{Same as in Table II, but for $95\%$ Confidence Level}
\begin{ruledtabular}
\begin{tabular}{cccc}
Luminosity($fb^{-1}$)&$\delta_{sys}$&$|a_{A}|$ & $a_{V}$  \\
\hline
$10$&$0\%$&$0.8894$ &$ (-1.2279;0.6453)$ \\
$10$&$3\%$&$1.0247$ &$ (-1.3574;0.7748)$ \\
$10$&$5\%$&$1.2616$ &$ (-1.5870;1.004)$ \\
\hline
$50$&$0\%$&$0.5948$ &$ (-0.9540;0.3714)$ \\
$50$&$3\%$&$0.8808$ &$ (-1.2197;0.6371)$ \\
$50$&$5\%$&$1.1940$ &$ (-1.5212;0.9386)$ \\
\hline
$100$&$0\%$&$0.5001$ &$ (-0.8704;0.2878)$ \\
$100$&$3\%$&$0.8570$ &$ (-1.1971;0.6144)$ \\
$100$&$5\%$&$1.1847$ &$ (-1.5122;0.9296)$ \\
\hline
$300$&$0\%$&$0.3800$ &$ (-0.7704;0.1877)$ \\
$300$&$3\%$&$0.8399$ &$ (-1.1809;0.5983)$ \\
$300$&$5\%$&$1.1783$ &$ (-1.5061;0.9234)$ \\
\hline
\end{tabular}
\end{ruledtabular}
\end{table}

\begin{table} [!htb]
\caption{The limits obtained at $68\%$ Confidence Level on the electromagnetic dipole moments of the top quark via the process $pp\rightarrow p \gamma^{*} p \rightarrow p t W X$ at the HL-LHC.}
\begin{ruledtabular}
\begin{tabular}{cccc}
Luminosity($fb^{-1}$)&$\delta_{sys}$&$|a_{A}|$ & $a_{V}$  \\
\hline
$500$&$0\%$&$0.2289$ &$ (-0.6682;0.0855)$ \\
$500$&$5\%$&$0.5974$ &$ (-0.9564;0.3738)$ \\
$500$&$10\%$&$0.8408$ &$ (-1.1818;0.5992)$ \\
\hline
$1500$&$0\%$&$0.1815$ &$ (-0.6246;0.0419)$ \\
$1500$&$5\%$&$0.5949$ &$ (-0.9541;0.3715)$ \\
$1500$&$10\%$&$0.8399$ &$ (-1.1810;0.5983)$ \\
\hline
$2500$&$0\%$&$0.1597$ &$ (-0.6236;0.0410)$ \\
$2500$&$5\%$&$0.5943$ &$ (-0.9536;0.3710)$ \\
$2500$&$10\%$&$0.8397$ &$ (-1.1808;0.5981)$ \\
\hline
$3000$&$0\%$&$0.1526$ &$ (-0.6202;0.0376)$ \\
$3000$&$5\%$&$0.5942$ &$ (-0.9535;0.3709)$ \\
$3000$&$10\%$&$0.8397$ &$ (-1.1807;0.5981)$ \\
\hline
\end{tabular}
\end{ruledtabular}
\end{table}

\begin{table} [!htb]
\caption{Same as in Table V, but for $90\%$ Confidence Level}
\begin{ruledtabular}
\begin{tabular}{cccc}
Luminosity($fb^{-1}$)&$\delta_{sys}$&$|a_{A}|$ & $a_{V}$  \\
\hline
$500$&$0\%$&$0.3065$ &$ (-0.7144;0.1317)$ \\
$500$&$5\%$&$0.7665$ &$ (-1.1119;0.5293)$ \\
$500$&$10\%$&$1.0788$ &$ (-1.4096;0.8270)$ \\
\hline
$1500$&$0\%$&$0.2329$ &$ (-0.6491;0.0665)$ \\
$1500$&$5\%$&$0.7633$ &$ (-1.1089;0.5262)$ \\
$1500$&$10\%$&$1.0777$ &$ (-1.4085;0.8259)$ \\
\hline
$2500$&$0\%$&$0.2050$ &$ (-0.6476;0.0650)$ \\
$2500$&$5\%$&$0.7642$ &$ (-1.1082;0.5256)$ \\
$2500$&$10\%$&$1.0774$ &$ (-1.4083;0.8256)$ \\
\hline
$3000$&$0\%$&$0.1958$ &$ (-0.6424;0.0598)$ \\
$3000$&$5\%$&$0.7624$ &$ (-1.1081;0.5254)$ \\
$3000$&$10\%$&$1.0774$ &$ (-1.4082;0.8256)$ \\
\hline
\end{tabular}
\end{ruledtabular}
\end{table}

\begin{table} [!htb]
\caption{Same as in Table V, but for $95\%$ Confidence Level}
\begin{ruledtabular}
\begin{tabular}{cccc}
Luminosity($fb^{-1}$)&$\delta_{sys}$&$|a_{A}|$ & $a_{V}$  \\
\hline
$500$&$0\%$&$0.3344$ &$ (-0.7350;0.1524)$ \\
$500$&$5\%$&$0.8363$ &$ (-1.1776;0.5949)$ \\
$500$&$10\%$&$1.1771$ &$ (-1.5048;0.9222)$ \\
\hline
$1500$&$0\%$&$0.2541$ &$ (-0.6780;0.0954)$ \\
$1500$&$5\%$&$0.8327$ &$ (-1.1742;0.5915)$ \\
$1500$&$10\%$&$1.1758$ &$ (-1.5036;0.9209)$ \\
\hline
$2500$&$0\%$&$0.2236$ &$ (-0.6587;0.0760)$ \\
$2500$&$5\%$&$0.8320$ &$ (-1.1735;0.5909)$ \\
$2500$&$10\%$&$1.1755$ &$ (-1.5033;0.9207)$ \\
\hline
$3000$&$0\%$&$0.2137$ &$ (-0.6527;0.0700)$ \\
$3000$&$5\%$&$0.8318$ &$ (-1.1733;0.5907)$ \\
$3000$&$10\%$&$1.1755$ &$ (-1.5033;0.9206)$ \\
\hline
\end{tabular}
\end{ruledtabular}
\end{table}

\begin{table} [!htb]
\caption{The limits obtained at $68\%$ Confidence Level on the electromagnetic dipole moments of the top quark via the process $pp\rightarrow p \gamma^{*} p \rightarrow p t W X$ at the HE-LHC.}
\begin{ruledtabular}
\begin{tabular}{cccc}
Luminosity($fb^{-1}$)&$\delta_{sys}$&$|a_{A}|$ & $a_{V}$  \\
\hline
$1000$&$0\%$&$0.1161$ &$ (-0.3481;0.0387)$ \\
$1000$&$5\%$&$0.4574$ &$ (-0.6375;0.3281)$ \\
$1000$&$10\%$&$0.6466$ &$ (-0.8194;0.5100)$ \\
\hline
$5000$&$0\%$&$0.0776$ &$ (-0.3278;0.0184)$ \\
$5000$&$5\%$&$0.4571$ &$ (-0.6372;0.3278)$ \\
$5000$&$10\%$&$0.6464$ &$ (-0.8193;0.5099)$ \\
\hline
$10000$&$0\%$&$0.0653$ &$ (-0.3226;0.0132)$ \\
$10000$&$5\%$&$0.4571$ &$ (-0.6372;0.3278)$ \\
$10000$&$10\%$&$0.6464$ &$ (-0.8193;0.5099)$ \\
\hline
$15000$&$0\%$&$0.0590$ &$ (-0.3202;0.0108)$ \\
$15000$&$5\%$&$0.4571$ &$ (-0.6372;0.3278)$ \\
$15000$&$10\%$&$0.6464$ &$ (-0.8193;0.5099)$ \\
\hline
\end{tabular}
\end{ruledtabular}
\end{table}

\begin{table} [!htb]
\caption{Same as in Table VIII, but for $90\%$ Confidence Level}
\begin{ruledtabular}
\begin{tabular}{cccc}
Luminosity($fb^{-1}$)&$\delta_{sys}$&$|a_{A}|$ & $a_{V}$  \\
\hline
$1000$&$0\%$&$0.1490$ &$ (-0.3694;0.0600)$ \\
$1000$&$5\%$&$0.5868$ &$ (-0.7615;0.4521)$ \\
$1000$&$10\%$&$0.8296$ &$ (-0.9985;0.6891)$ \\
\hline
$5000$&$0\%$&$0.0996$ &$ (-0.3387;0.0293)$ \\
$5000$&$5\%$&$0.5865$ &$ (-0.7612;0.4518)$ \\
$5000$&$10\%$&$0.8294$ &$ (-0.9983;0.6889)$ \\
\hline
$10000$&$0\%$&$0.0838$ &$ (-0.3306;0.0212)$ \\
$10000$&$5\%$&$0.5865$ &$ (-0.7612;0.4518)$ \\
$10000$&$10\%$&$0.8294$ &$ (-0.9983;0.6889)$ \\
\hline
$15000$&$0\%$&$0.0757$ &$ (-0.3269;0.0175)$ \\
$15000$&$5\%$&$0.5865$ &$ (-0.7612;0.4518)$ \\
$15000$&$10\%$&$0.8294$ &$ (-0.9983;0.6889)$ \\
\hline
\end{tabular}
\end{ruledtabular}
\end{table}

\begin{table} [!htb]
\caption{Same as in Table VIII, but for $95\%$ Confidence Level}
\begin{ruledtabular}
\begin{tabular}{cccc}
Luminosity($fb^{-1}$)&$\delta_{sys}$&$|a_{A}|$ & $a_{V}$  \\
\hline
$1000$&$0\%$&$0.1625$ &$ (-0.3791;0.0697)$ \\
$1000$&$5\%$&$0.6403$ &$ (-0.8133;0.5039)$ \\
$1000$&$10\%$&$0.9051$ &$ (-1.0728;0.7634)$ \\
\hline
$5000$&$0\%$&$0.1087$ &$ (-0.3437;0.0343)$ \\
$5000$&$5\%$&$0.6399$ &$ (-0.8129;0.5035)$ \\
$5000$&$10\%$&$0.9049$ &$ (-1.0726;0.7632)$ \\
\hline
$10000$&$0\%$&$0.0914$ &$ (-0.3343;0.0249)$ \\
$10000$&$5\%$&$0.6399$ &$ (-0.8129;0.5035)$ \\
$10000$&$10\%$&$0.9049$ &$ (-1.0726;0.7632)$ \\
\hline
$15000$&$0\%$&$0.0826$ &$ (-0.3300;0.0206)$ \\
$15000$&$5\%$&$0.6399$ &$ (-0.8129;0.5035)$ \\
$15000$&$10\%$&$0.9049$ &$ (-1.0726;0.7632)$ \\
\hline
\end{tabular}
\end{ruledtabular}
\end{table}

However, we compare our results with the limits derived at the LHC and beyond as examined in Refs. \cite{1,2,4,mur}. As can be seen in Table VII, the sensitivities of the anomalous couplings via the process $pp\rightarrow p \gamma^{*} p \rightarrow p t W X$ at the HL-LHC with $L_{int}=3$ ab$^{-1}$ are at the same order with those reported in Refs. \cite{1,2}. In Ref. \cite{4}, the best limits on $a_{A}$ and $a_{V}$ couplings by probing the process $pp \to p\gamma^*\gamma^*p\to pt\bar t p $ at LHC-33 TeV with $L_{int}=3000$ fb$^{-1}$ are found. We see from Table VIII that our limits obtained on $a_{A}$ and $a_{V}$ couplings for $L_{int}=15000$ fb$^{-1}$ are nearly 1.5 times better than those reported in Ref. \cite{4}. In Ref. \cite{mur}, the best sensitivities derived from the process $pp\rightarrow p \gamma^{*} p \rightarrow p t \bar{t} X$ at the HE-LHC are obtained as $|a_{A}|=0.020$ and $a_{V}=[-0.9953;0.0003]$. On the other hand, we understand that the sensitivities on the anomalous $a_{A}$ and $a_{V}$ couplings expected to be obtained for the examined process with at the HE-LHC with $L_{int}=15$ ab$^{-1}$ as shown in Table VIII are more worse than the sensitivities in Ref. \cite{mur}. This is because the distribution function of gluon in the proton is greater than the distribution function of the bottom quark.

We take into account on the limits of the anomalous couplings including systematic errors. The best limits calculated with without systematic error for $a_{A}$ coupling are nearly an order of magnitude better than the results of $10\%$ systematic error. Also, we see that the limits derived on the positive part of $a_{V}$ coupling with without systematic error can set more stringent sensitive with respect  to the results obtained on the negative part of $a_{V}$ coupling. The reason for these behaviors can be easily understood from Figs. 3 and 4. As can be understood from these figures, deviation from the SM cross section of the positive part of $a_{V}$ coupling is greater than the negative part of $a_{V}$ coupling.

The two dimensional limits at $95\%$ Confidence Level intervals in planes of the anomalous $a_{A}$ and $a_{V}$ couplings are presented in Figs. 7-9.  As we understand from Figs. 7-9, the improvement in the limit on the anomalous parameters is reached by increasing to higher luminosities and center-of-mass energies.

\begin{figure}
\includegraphics{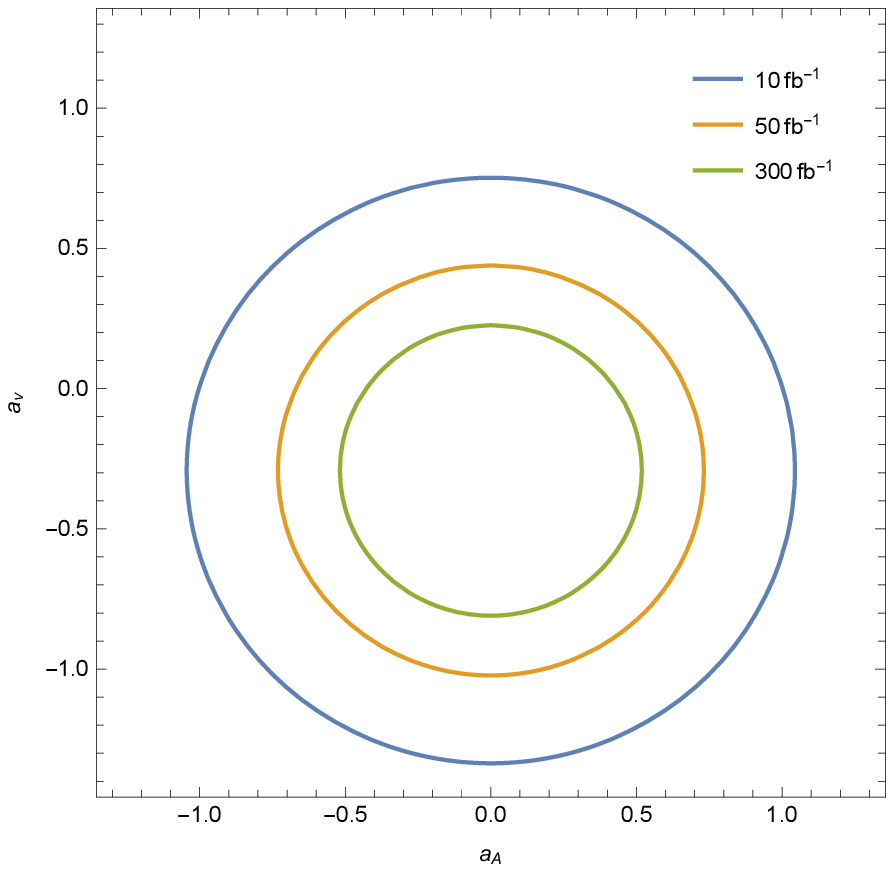}
\caption{Contours for the anomalous $a_{A}$ and $a_{V}$ couplings for the process $pp\rightarrow p \gamma^{*} p \rightarrow p t W X$ at the LHC.
\label{fig7}}
\end{figure}

\begin{figure}
\includegraphics{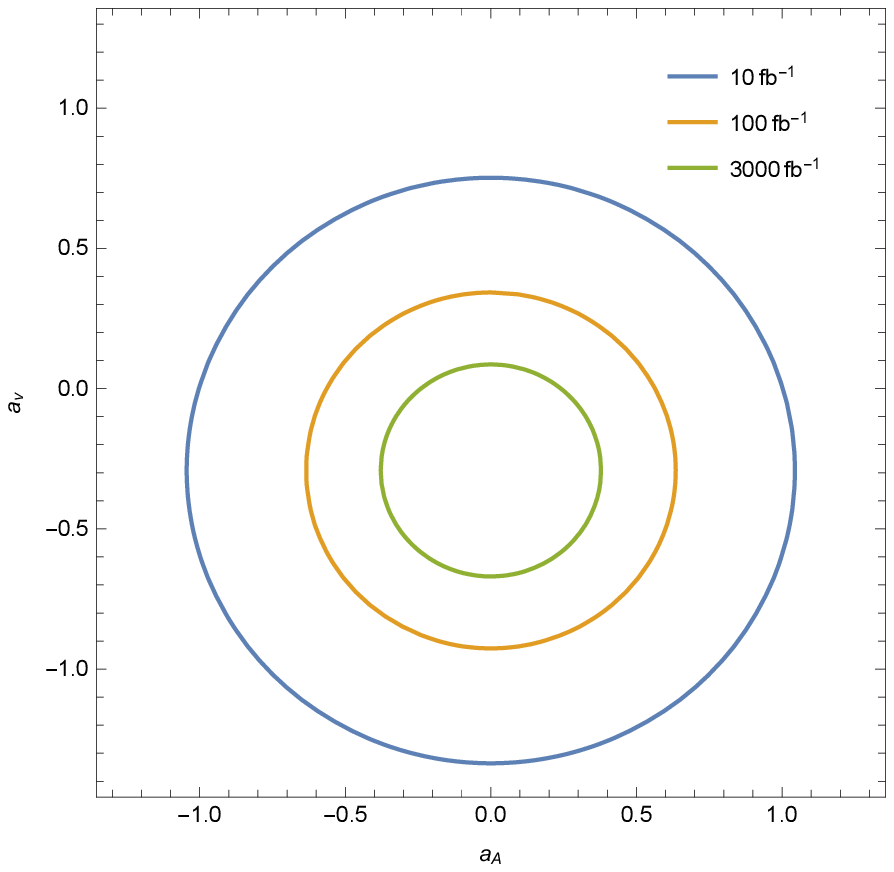}
\caption{Same as in Fig. 7, but for the HL-LHC.
\label{fig8}}
\end{figure}

\begin{figure}
\includegraphics{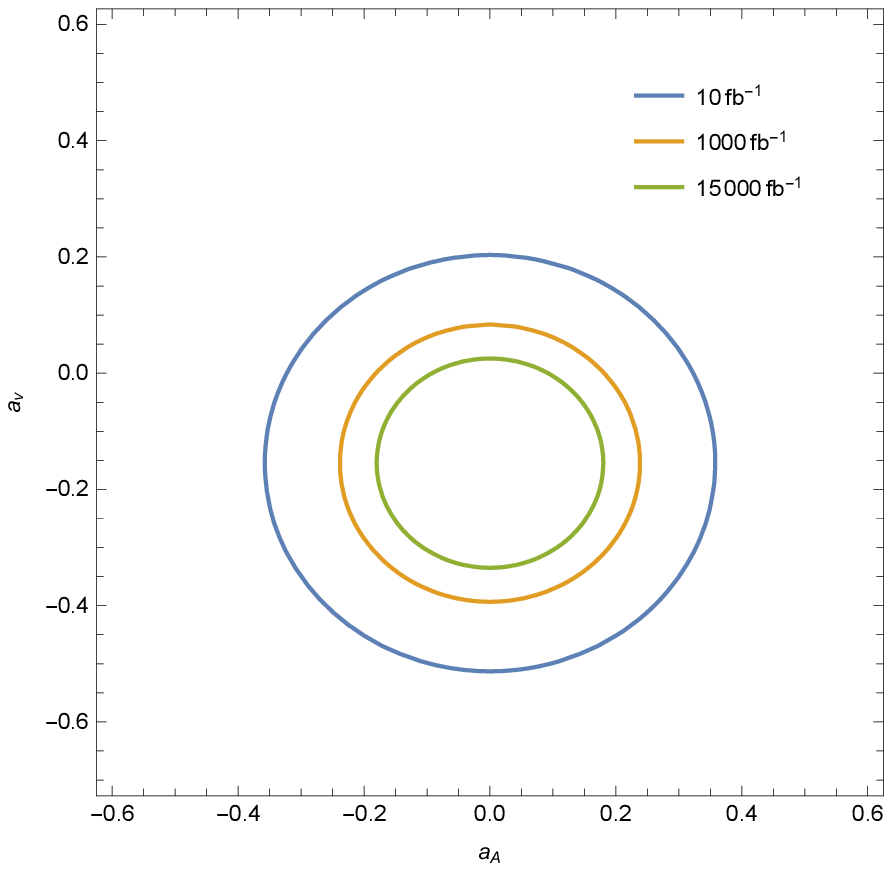}
\caption{Same as in Fig. 7, but for the HE-LHC.
\label{fig9}}
\end{figure}

\section{Conclusions}

The upgrade of the LHC to a HL-LHC at $\sqrt{s}=14$ TeV with $L_{int}=3$ ab$^{-1}$ will examine the SM with even greater precision and will enlarge the sensitivity to likely anomalies in the SM. For the HE-LHC investigates the dataset is supposed to be $L_{int}=15$ ab$^{-1}$ at $\sqrt{s}=27$ TeV. It is clear that the HE-LHC is extremely important to produce the heaviest particles.

Top quarks at the LHC are copiously generated in collisions, providing a rich testing ground for theoretical models of particle collisions at the highest accessible energies. Since the anomalous $t\bar{t}\gamma$ couplings defined through effective Lagrangian that has dimension-6, they have very strong energy dependencies. For this reason, the total cross section of the subprocess $\gamma^{*} b \rightarrow t W$ including the anomalous $t\bar{t}\gamma$ coupling has higher energy than the SM cross section. Thus, any deviations between measurements and predictions could point to shortcomings in the theory or first hints of something completely new. The second important point is that research of the anomalous $t\bar{t}\gamma$ couplings presents one of the significant alternatives to find out new physics such as the electric dipole moment of the top quark is particularly attractive since it is very sensitive to possible new sources of CP violation in the lepton and quark sectors. Hence, it is very important to examine the anomalous $t\bar{t}\gamma$ couplings at LHC and beyond.

The photon-induced $\gamma^{*} \gamma^{*}$ and $\gamma^{*} p$ collisions ensure an increase of the physics examined at $pp$ colliders.
However, $\gamma^{*} p$ collisions have high center-of-mass energy and high luminosity with respect to $\gamma^{*} \gamma^{*}$ collisions.
In addition, $\gamma^{*} p$ collisions due to the remnants of only one of the proton beams have fewer backgrounds than $pp$ collisions. Therefore,
we expect that $\gamma^{*} p$ collisions have a higher potential in examining new physics beyond the SM compared with $\gamma^{*} \gamma^{*}$ collisions.

Therefore, in this study, we study the potential of the process $pp\rightarrow p \gamma^{*} p \rightarrow p t W X$ at the LHC, the HL-LHC and the HE-LHC to investigate the electromagnetic dipole moments of the top quark. We obtain limits at $68\%$, $90\%$ and $95\%$ Confidence Levels on the anomalous $a_{A}$ and $a_{V}$ couplings with various values of the center-of-mass energy and the integrated luminosity.  

As a result, the LHC and beyond as a $\gamma^{*} p$ collisions provide us a good opportunity to probe the electric and magnetic dipole moments of the top quark.

\pagebreak

\newpage


\begin{thebibliography}{99}

\bibitem{cp1} J. H. Christenson, J. W. Cronin, V. L. Fitch and R. Turlay, Phys. Rev. Lett. 13, 138 (1964).
\bibitem{cp2} M. Kobayashi and T. Maskawa, Prog. Teor. Phys. 49, 652 (1973).
\bibitem{yam} N Yamanaka, T. Yamada and Y. Funaki,  arXiv:1907.08091.
\bibitem{yam1} K. Yanase, N. Yoshinaga, K. Higashiyama and N. Yamanaka, Phys. Rev. D 99, no.7, 075021 (2019).
\bibitem{yam2} N. Yamanaka,  Int. J. Mod. Phys. E 26, no.4, 1730002 (2017).
\bibitem{cp3} D. Atwood, S. Bar-Shalom, G. Eilam and A. Soni, Phys. Rept. 347, 1 (2001).
\bibitem{cp4} A. Brandenburg and J. P. Ma, Phys. Lett. B 298, 211 (1993).
\bibitem{sm} F. Hoogeveen, Nucl. Phys. B 341, 322 (1990).
\bibitem{sm1} W. Bernreuther, R. Bonciani, T. Gehrmann, R. Heinesch, T. Leineweber, P. Mastrolia and E. Remiddi,
Phys. Rev. Lett. 95, 261802 (2005).
\bibitem{1} U. Baur, A. Juste, L. H. Orr and D. Rainwater, Phys. Rev. D 71, 054013 (2005).
\bibitem{2} S. M. Etesami, S. Khatibi and M. M. Najafabadi, Eur. Phys. J. C 76, 533 (2016).
\bibitem{4} Sh. Fayazbakhsh, S. Taheri Monfared and M. Mohammadi Najafabadi, Phys. Rev. D 92, 014006 (2015).
\bibitem{mur} M. Koksal, arXiv:1906.09287.
\bibitem{5} Aguilar-Saavedra J. A., {\it et al.}, [ECFA/DESY LC Physics Working Group Collaboration], hep-ph/0106315.
\bibitem{6} M. Koksal, A. A. Billur and A. Gutierrez-Rodriguez, Adv. High Energy Phys. 2017, 6738409 (2017).
\bibitem{7} A. A. Billur, M. Koksal and A. Gutierrez-Rodriguez, Phys. Rev. D 96, 056007 (2017).
\bibitem{8} M. Koksal, A.A. Billur, A. Gutierrez-Rodriguez and M. A. Hernandez-Ruiz, arXiv:1905.02564.
\bibitem{tur} M. A. Hernandez-Ruiz, A. Gutierrez-Rodriguez, M. Koksal and A. A. Billur, Nucl. Phys. B 941, 646-664 (2019).
\bibitem{10} A. A. Billur, M. Koksal, A. Gutierrez-Rodriguez and M. A. Hernandez-Ruiz, arXiv:1811.10462.
\bibitem{9} A. O. Bouzas and F. Larios, Phys. Rev. D 88, 0094007 (2013).
\bibitem{20} A. O. Bouzas and F. Larios, Phys. Rev. D 87, 074015 (2013).
\bibitem{21} K. Fuyuto and M. Ramsey-Musolf, Phys. Lett. B 781, 492-498 (2018).
\bibitem{22} V. Cirigliano, W. Dekens, J. de Vries, and E. Mereghetti, Phys. Rev. D 94, 016002 (2016).
\bibitem{23} V. Cirigliano, W. Dekens, J. de Vries, and E. Mereghetti, Phys. Rev. D 94, 034031 (2016).
\bibitem{24} J. Montano, H. Novales-Sanchez and J. J. Toscano, arXiv:1908.06226.
\bibitem{3} M. Fael and T. Gehrmann, Phys. Rev. D 88, 033003 (2013).
\bibitem{HL} High-Luminosity Large Hadron Collider (HL-LHC): Technical Design Report V. 0.1. CERN Yellow Rep.Monogr. 4, 1-516 (2017).
\bibitem{HE} FCC Collaboration, Eur. Phys. J. ST 228, no.5, 1109-1382 (2019).
\bibitem{ckm} S. Ovyn, Nucl. Phys. B 179-180, 269-276 (2008).
\bibitem{ckm2} B. Sahin and A. A. Billur, Phys. Rev. D 86, 074026 (2012).
\bibitem{ahe} H. Sun, Nucl.Phys. B 886, 691-711 (2014).
\bibitem{ah1} R. Goldouzian and B. Clerbaux, Phys. Rev. D 95, 054014 (2017).
\bibitem{ahe1} H. Sun, Eur. Phys. J. C 74, 2823 (2014).
\bibitem{ahe2} J. de Favereau de Jeneret \textit{et al}., arXiv:0908.2020.
\bibitem{epa} V. M. Budnev, I. F. Ginzburg, G. V. Meledin and V. G. Serbo, Phys. Rep. 15, 181 (1975).
\bibitem{epa1} G. Baur \textit{et al}., Phys. Rep. 364, 359 (2002).
\bibitem{epa2} K. Piotrzkowski, Phys. Rev. D 63, 071502 (2001).
\bibitem{s1} I. Sahin and M. Koksal, JHEP 11, 100 (2011).
\bibitem{s2} I. Sahin and A. A. Billur, Phys. Rev. D 83, 035011 (2011).
\bibitem{s3} S. C. Inan and A. V. Kisselev, arXiv::1902.08615.
\bibitem{s4} S. C. Inan and A. V. Kisselev, Eur. Phys. J. C 78, no.9, 729 (2018).
\bibitem{s5} M. Koksal \textit{et al}., Phys. Lett. B 783, 375-380 (2018).
\bibitem{s6} A. A. Billur, Europhys. Lett. 101, 21001 (2013).
\bibitem{s7} M. Koksal and S. C. Inan, Adv. High Energy Phys. 2014, 935840 (2014).
\bibitem{s8} M. Koksal and S. C. Inan, Adv. High Energy Phys. 2014, 315826 (2014).
\bibitem{s9} S. C. Inan, Nucl. Phys. B 897, 289(2015).
\bibitem{s10} P. Xue-An \textit{et al}., Phys. Rev. D 99, 014029 (2019).
\bibitem{s11} A. Senol and M. Koksal, Phys. Lett. B 742, 143 (2015).
\bibitem{s12} D. Alva, T. Han and R. Ruiz, JHEP 1502, 072 (2015).
\bibitem{s13} S. C. Inan and A. A. Billur, Phys.Rev. D 84, 095002 (2011).
\bibitem{s15} B Sahin, Mod.Phys.Lett. A 32, no.37, 1750205 (2017).
\bibitem{s16} B Sahin, Adv. High Energy Phys. 2015, 590397 (2015).
\bibitem{s18} A. Esmaili, S. Khatibi and M. M. Najafabadi, Phys. Rev. D 96, 015027 (2017).
\bibitem{s19} A. Senol, Phys. Rev. D 87 073003 (2013).
\bibitem{s20} C. Baldenegro, S. Fichet, G. von Gersdorff, C. Royon,  JHEP 1706, 142 (2017).
\bibitem{s21} C. Baldenegro, S. Fichet, G. von Gersdorff, C. Royon, JHEP 1806, 131 (2018).
\bibitem{s22} S. Fichet, G. V. Gersdorff, B. Lenzi, C. Royon, M. Saimpert, JHEP 1502, 165 (2015).
\bibitem{sa1} G. Akkaya Selcin and \.{I}. \c{S}ahin, Chin. J. Phys. 55, 2305-2317 (2017).
\bibitem{sa2} I. Sahin \textit{et al}., Phys. Rev. D 91, 035017 (2015).
\bibitem{sa3} I. Sahin \textit{et al}., Phys. Rev. D 88,  095016 (2013).
\bibitem{sa4} I. Sahin and B. \c{S}ahin,  Phys.Rev. D 86, 115001 (2012).
\bibitem{sa5} I. Sahin, Phys. Rev. D 85, 033002 (2012).
\bibitem{sa6} H. Sun \textit{et al}., JHEP 1502, 064 (2015).
\bibitem{sa7} H. Sun, Phys. Rev. D 90, 035018 (2014).
\bibitem{sa10} A. Senol, Int. J. Mod. Phys. A 29, 1450148 (2014).
\bibitem{sa11} A. Senol \textit{et al}., Mod. Phys. Lett. A 29  no.36, 1450186.
\bibitem{sa12} G-C. Cho, K. Yamashita and M. Yonemura, arXiv:1908.06357 .
\bibitem{Kamenik} J. F. Kamenik, M. Papucci and A. Weiler, Phys. Rev. D 85, 071501 (2012).
\bibitem{Baur} U. Baur, A. Juste, L. H. Orr and D. Rainwater, Phys. Rev. D 71, 054013 (2005).
\bibitem{Aguilar1} J. A. Aguilar-Saavedra, M. C. N. Fiolhais and A. Onofre, JHEP 07, 180 (2012).
\bibitem{calc} A. Belyaev, N. D. Christensen and A. Pukhov, Comput. Phys. Commun. 184, 1729 (2013).
\bibitem{cte} J. Pumplin, D. R. Stump, J. Huston, H. L. Lai, P. Nadolsky and W. K. Tung, JHEP 07, 012
(2002).
\bibitem{top} CMS Collaboration, JHEP 10, 117 (2018).


\end{thebibliography}
\end{document}